\documentclass[12pt, centerh1]{article}

\usepackage{graphicx,colonequals,bm}
\usepackage{natbib}
\usepackage{url}
\usepackage{amsmath}
\usepackage{amsfonts}
\usepackage{amssymb}
\usepackage{url}
\usepackage{color}
\usepackage{subfigure}
\usepackage{slashbox}
\newcommand{\be}{\begin{equation}}
\newcommand{\ee}{\end{equation}} 
\newcommand{\beq}{\begin{equation*}}
\newcommand{\eeq}{\end{equation*}}

\newcommand{\diag}{\,\mbox{diag}}

\newcommand{\load}{\mathbf\Lambda}

\newcommand{\Beta}{\boldsymbol\beta}
\newcommand{\sampcov}{\mathbf{S}}
\newcommand{\ident}{\mathbf{I}}
\newcommand{\vecA}{\mathbf{A}}
\newcommand{\vecC}{\mathbf{C}}
\newcommand{\vecD}{\mathbf{D}}
\newcommand{\vecY}{\mathbf{Y}}
\newcommand{\vecy}{\mathbf{y}}
\newcommand{\vecx}{\mathbf{x}}
\newcommand{\vecX}{\mathbf{X}}
\newcommand{\vecU}{\mathbf{U}}
\newcommand{\vecu}{\mathbf{u}}

\newcommand{\vecV}{\mathbf{V}}
\newcommand{\vecz}{\mathbf{z}}
\newcommand{\vecZ}{\mathbf{Z}}

\newcommand{\vecN}{\mathbf{N}}

\newcommand{\Ewg}{\mathbb{E}\left[ W_{ig} \mid \vecx_i, z_{ig}=1 \right]}

\newcommand{\Ewinvg }{\mathbb{E}[ W_{ig}^{-1} \mid \vecx_i, z_{ig}=1]}

\newcommand{\vecPsi}{\mbox{\boldmath$\Psi$}}

\newcommand{\varthet}{\mbox{\boldmath$\vartheta$}}
\newcommand{\vectheta}{\mbox{\boldmath$\theta$}}
\newcommand{\del}{\mbox{\boldmath$\Delta$}}
\newcommand{\vecmu}{\mbox{\boldmath$\mu$}}

\newcommand{\vecalpha}{\mbox{\boldmath$\alpha$}}

\newcommand{\mSigma}{\mbox{\boldmath$\Sigma$}}
\newcommand{\matsig}{\mSigma}
\newcommand{\vecLambda}{\mbox{\boldmath$\Lambda$}}
\newcommand{\vecepsilon}{\mbox{\boldmath$\epsilon$}}

\newcommand{\lsum}{\sum_{i=1}^n}
\newcommand{\lssum}{\sum_{j=1}^n}
\newcommand{\gsum}{\sum_{g=1}^G}

\newcommand{\zig}{\hat{z}_{ig}}
\newcommand{\zjg}{\hat{z}_{jg}}

\begin{document}

\title{Parsimonious Shifted Asymmetric Laplace Mixtures}
\author{Brian C.\ Franczak, Paul D.\ McNicholas\thanks{Department of Mathematics \& Statistics, University of Guelph, Guelph, Ontario, N1G 2W1, Canada. E-mail: pmcnicho@uoguelph.ca.}, Ryan P.\ Browne and Paula M.\ Murray}
\date{Department of Mathematics \& Statistics, University of Guelph}
\maketitle

\begin{abstract}
A family of parsimonious shifted asymmetric Laplace mixture models is introduced. We extend the mixture of factor analyzers model to the shifted asymmetric Laplace distribution. Imposing constraints on the constitute parts of the resulting decomposed component scale matrices leads to a family of parsimonious models. An explicit two-stage parameter estimation procedure is described, and the Bayesian information criterion and the integrated completed likelihood are compared for model selection. This novel family of models is applied to real data, where it is compared to its Gaussian analogue within clustering and classification paradigms.
%
\end{abstract}

\section{Introduction}\label{sec:introduction}

Model-based clustering is an umbrella term that describes fitting finite mixture models to data for clustering, i.e., to recover underlying subpopulations. This approach can be traced back to \cite{wolfe63} and due to its natural appeal, it has become more and more popular in recent years \citep[e.g.,][]{mcnicholas08,karlis09,andrews11d,browne12}.
A random variable $\vecX$ from a finite mixture model has density of the form 
\beq
f(\vecx\mid\varthet)=\gsum{\pi_gf_g(\vecx\mid\vectheta_g)},
\eeq
where $\pi_g>0$ are the mixing proportions with $\gsum{\pi_g} = 1$, $f_1(\vecx\mid\vectheta_g),\ldots,f_G(\vecx\mid\vectheta_G)$ are the component densities, and $\varthet=(\pi_1,\dots\pi_G,\vectheta_1,\dots,\vectheta_G)$ denotes the model parameters. 
Although the component densities are free to take on many forms, they are typically taken to be of the same type, most often multivariate Gaussian \citep[e.g.,][]{banfield93,celeux95,ghahramani97,mclachlan00a,mclachlan03,mcnicholas10a}. 
Formally, a random vector $\vecX$ is said to arise from a finite mixture of multivariate Gaussian distributions if its density has the form
\be\label{eq:gaussmix1}
f(\vecx\mid\varthet) = \gsum{\pi_g\phi(\vecx\mid\vecmu_g,\matsig_g)},
\ee
where $\phi(\vecx\mid\vecmu_g,\mSigma_g)$ is the density of the multivariate Gaussian distribution with  mean $\vecmu_g$ and covariance matrix $\mSigma_g$. 

To increase parsimony and flexibility, families of finite mixture models have been developed by imposing a combination of constraints on the component parameters, most commonly the component covariance matrices \citep[e.g.,][]{celeux95,andrews12,mcnicholas12,browne12,subedi13}. The most well-known family of mixture models is the Gaussian parsimonious clustering models (GPCM) family \citep{banfield93,celeux95}.

The GPCMs result from constraining the elements of the eigen-decomposed component covariance matrices, i.e., $\mSigma_g = \lambda_g\vecD_g\vecA_g\vecD_g'$, by imposing combinations of the constraints $\vecD_g=\vecD$, $\vecD=\ident_d$, $\vecA_g=\vecA$, $\vecA_g=\ident_d$, and $\lambda_g=\lambda$.  In this decomposition, $\lambda_g$ is a constant, $\vecD_g$ is a $p\times p$ matrix of eigenvectors of $\matsig_g$, and $\vecA_g$ is a $p\times p$ diagonal matrix with entries proportional to the eigenvectors of $\matsig_g$ and $|\vecA_g| = 1$. In total, there are fourteen models in the GPCM family. All 14 are available in the \textsf{R} \citep{R12} packages \texttt{Rmixmod} \citep{lebret12} and \texttt{mixture} \citep{browne12,browne13,browne13a}, and 10 are available in the \textsf{R} package \texttt{mclust} \citep{fraley02a}.

More recently, non-Gaussian analogues of the GPCM family have been developed. \cite{andrews12} introduce a $t$-analogue, which is available as the \texttt{teigen} package \citep{andrews13} for \textsf{R}. \cite{vrbik13} introduce skew-normal and skew-t analogues of the GPCM family. These families offer robust and flexible alternatives compared to the well-established family of Gaussian mixture models. Although some of the GPCM models significantly reduce the number of free parameters in the component covariance matrices, eight of the 14 still have a number of component covariance parameters that are quadratic in $p$. Of course, the same will hold for the component scale matrices in non-Gaussian analogues of the GPCM family. 

An alternative approach to introducing parsimony is to assume underlying latent variables of (much) lower dimension. The mixture of factor analyzers model \citep{ghahramani97,mclachlan00a} is the most popular such approach within the literature. Factor analysis \citep{spearman04} is a dimension reduction technique that assumes a $p$-dimensional random vector $\vecX$ can be modelled using a $q$-dimensional vector of unobserved (or latent) factors $\vecU$, where $q\ll p$. We can write the factor analysis model as \begin{equation*}\vecX = \vecmu + \vecLambda\vecU+\vecepsilon,\end{equation*} where $\vecLambda$ is a $p\times q$ matrix of factor loadings, $\vecU\backsim\mathcal{N}(\bm{0},\ident_q)$ is a vector of factors, and $\vecepsilon\backsim\mathcal{N}(\bm{0},\vecPsi)$ is a vector of error terms with $\vecPsi = \diag{(\psi_1,\psi_2,\dots,\psi_p)}$. It follows that the marginal distribution of $\vecX$ is 
multivariate Gaussian with mean $\vecmu$ and covariance matrix $\mSigma = \vecLambda\vecLambda'+\vecPsi$.

The mixture of factor analyzers model is given by a Gaussian mixture with component covariance matrices \begin{equation*}\mSigma_g = \vecLambda_g\vecLambda_g'+\vecPsi_g,\end{equation*} where $\vecLambda_g$ is a $p\times q$ matrix of factor loadings, and $\vecPsi_g$ is a $p\times p$ diagonal matrix with positive entries. The closely related (i.e., $\vecPsi_g=\psi_g\ident_p$) mixture of probabilistic principal component analyzers model was studied by \cite{tipping99b}, and \cite{mcnicholas08} extended the mixture of factor analyzers model to a family of eight parsimonious models. 
\cite{mcnicholas10d} modified the factor analysis component covariance structure by setting $\vecPsi_g=\omega_g\del_g$, where $\del_g$ is a $p\times p$ diagonal matrix with $|\del_g|=1$, and $\omega_g\in\mathbb{R}^{+}$. The mixture of modified factor analyzers model is given by a Gaussian mixture with component covariance matrices \begin{equation*}\mSigma_g = \vecLambda_g\vecLambda_g'+\omega_g\del_g.\end{equation*} By imposing valid combinations of the constraints $\vecLambda_g = \vecLambda$, $\del_g = \ident_p$, $\del_g=\del$, and $\omega_g=\omega$, a family of 12 mixtures of modified factor analyzers emerges \citep[cf.][]{mcnicholas10d}. This family  is referred to as the parsimonious Gaussian mixture models (PGMM) family, and is available as the \textsf{R} package \texttt{pgmm} \citep{mcnicholas11}.

In this paper, we extend the mixture of factor analyzers using a mixture of shifted asymmetric Laplace (SAL) distributions \citep{franczak13}. Then, a SAL-analogue of the PGMM family is developed. The rest of this paper is laid out as follows. Some background material is presented in Section~\ref{sec:back}. Then our methodology is discussed as well as parameter estimation (Section~\ref{sec:meth}). Computational issues are discussed (Section~\ref{sec:CC}) and our approach is illustrated on real data (Section~\ref{sec:Apps}). The paper concludes with discussion and suggestions for future work (Section~\ref{sec:discuss}).

\section{Background}\label{sec:back}

\subsection{Shifted Asymmetric Laplace Distributions}\label{sec:sal}

\cite{franczak13} introduce a mixture of SAL distributions. The density of a $g$-component SAL mixture model is given by \begin{equation*}
f(\vecx\mid\varthet)=\sum_{g=1}^G\pi_g\xi\left(\vecx\mid\vectheta_g\right),
\end{equation*} where
\be\begin{split}\label{eq:SAL}
&\xi\left(\vecx\mid\vectheta_g\right)=
\frac{2\exp\{(\vecx-\vecmu_g)'\matsig_g^{-1}\vecalpha_g\}}
{(2\pi)^{p/2}\vert\matsig_g\vert^{1/2}}\left[\frac{\delta_g\left(\vecx,\vecmu_g\mid\matsig_g\right)}
{2+\vecalpha_g'\matsig_g^{-1}\vecalpha_g}\right]^{\nu_g/2}\\
&\qquad\qquad\qquad\qquad\qquad\qquad\qquad\qquad\times
K_{\nu}\left(\sqrt{(2+\vecalpha_g'\matsig_g^{-1}\vecalpha_g)\delta\left(\vecx,\vecmu_g\mid\matsig_g\right)}\right),
\end{split}\ee
$\nu=(2-p)/2$, $\mSigma_g$ is a scale matrix, $\vecalpha_g\in\mathbb{R}^p$ is a skewness parameter, $\vecmu_g\in\mathbb{R}^p$ is a location parameter, $K_{\nu}$ is the modified Bessel function of the third kind with index~$\nu$, $\delta\left(\vecx,\vecmu_g\mid\matsig_g\right)=\left(\vecx-\vecmu_g\right)'\matsig^{-1}\left(\vecx-\vecmu_g\right)$ is the squared Mahalanobis distance between $\vecx$ and $\vecmu_g$, $\vectheta_g=(\vecalpha_g, \matsig_g,\vecmu_g)$, and $\varthet$ denotes all model parameters. The SAL density given in (\ref{eq:SAL}) modifies the asymmetric Laplace density \citep{kotz01} via the addition of a shift parameter to facilitate clustering \citep[cf.][]{franczak13}. 

\subsection{Generalized Inverse Gaussian}\label{sec:GIG}

Let $X\backsim\text{GIG}(a,b)$ denote that a random variable $X$ follows the generalized inverse Gaussian (GIG) distribution with density,
\be\label{gig}
q(x) = \frac{(a/b)^{\nu/2}x^{\nu-1}}{2K_{\nu}(\sqrt{ab})} \exp\left\{-\frac{ax+b/x}{2}\right\},
\ee
for $x>0$, where $a,b\in\mathbb{R}^+$, $\nu\in\mathbb{R}$, and $K_{\nu}$ is defined as before. The GIG distribution has many nice properties \citep[cf.][]{barndorff77,blaesild78,halgreen79,jorgensen82}. For our purposes, the most enticing of these properties is the tractability of the following expected values:
\begin{equation}\begin{split}\label{eqn:exp_vals}
&\mathbb{E}\left[ X \right] =
\sqrt{\frac{b}{a}} R_{\nu}\left( \sqrt{ab}\right),\\
&\mathbb{E}\left[{1}/{X}\right] = 
\sqrt{\frac{a}{b}} R_{\nu}\left( \sqrt{ab}\right) -\frac{2\nu}{b},
\end{split}\end{equation}
where $R_{\nu}(z) \colonequals  K_{\nu+1}\left( z\right)/ K_{\nu}\left( z\right)$.

\subsection{Generalized Hyperbolic Distribution}\label{sec:GHD}

A random variable $\vecY$ following the generalized hyperbolic distribution has density,
\begin{equation}\begin{split} \label{eq:GHD}
&f(\vecy\mid\lambda,\chi,\psi,\vecmu,\matsig,\vecalpha) = 
\left[ \frac{ \chi + \delta\left(\vecy, \vecmu\mid\matsig\right) }{ \psi+ \vecalpha'\del^{-1}\vecalpha} \right]^{(\lambda-{p}/{2})/2}\\
&\qquad\qquad\qquad\qquad\qquad\qquad\times\frac{[{\psi}/{\chi}]^{{\lambda}/{2}}K_{\lambda - {p}/{2}}\left(\sqrt{[\psi+ \vecalpha'\matsig^{-1}\vecalpha] [ \chi +\delta(\vecy, \vecmu\mid\del)]}\right)}{ \left(2\pi\right)^{{p}/{2}} \left|\matsig\right|^{{1}/{2}} K_{\lambda}\left( \sqrt{\chi\psi}\right)\exp\left\{ \left( \vecmu- \vecy\right)'\matsig^{-1}\vecalpha\right\}},
\end{split}\end{equation}
where $\delta\left(\vecx,\vecmu\mid\matsig\right)$ is as defined before. \cite{mcneil05} discuss the limiting cases of the generalized hyperbolic distribution and show that if $\lambda = 1$, $\psi = 2$, and $\chi\rightarrow 0$ we obtain the multivariate asymmetric Laplace distribution \citep[cf.][]{kotz01}.

\section{Methodology}\label{sec:meth}

\subsection{The Model}\label{sec:model}

\cite{kotz01} note that a random variable $\vecV$ following the asymmetric Laplace distribution with skewness $\vecalpha$ and scale matrix $\mSigma$ can be generated through the relationship
\begin{equation*}
\vecV = W\vecalpha + \sqrt{W}\vecN,
\end{equation*}
where $W\backsim \text{Exp}(1)$ and $\vecN\backsim\mathcal{N}(\mathbf{0},\mSigma)$ are independent of one another. 
Adapting this relationship to  a random variable $\vecX\backsim\text{SAL}(\vecmu,\mSigma,\vecalpha)$ gives
\beq
\vecX = \vecmu+W\vecalpha+\sqrt{W}\vecN,
\eeq 
and it follows that $\vecX\mid w\backsim\mathcal{N}(\vecmu+w\vecalpha,w\mSigma)$. 
From Bayes' theorem, we have
\begin{equation}\begin{split}\label{densw}
f_W(w&\mid\vecX =\vecx)
=\frac{w^{\nu-1}}{2}
\left(\frac{\delta\left(\vecx,\vecmu\mid\matsig\right)}{2+\vecalpha'\matsig^{-1}\vecalpha}\right)^{-\nu/2}\frac{\exp\left\{-\frac{1}{2w}\delta\left(\vecx,\vecmu\mid\matsig\right)-\frac{w}{2}\left(2+\vecalpha'\matsig^{-1}\vecalpha\right)\right\}}{K_{\nu}\left(\sqrt{(2+\vecalpha'\matsig^{-1}\vecalpha)\delta(\vecx,\vecmu\mid\matsig)}\right)},
\end{split}\end{equation}
where $\nu$, $\vecalpha$, $\vecmu$, $\matsig$, and $\delta\left(\vecx,\vecmu\mid\matsig\right)$ are as defined for~\eqref{eq:SAL}. From \eqref{densw}, we have $W\mid\vecx\backsim\text{GIG}{(a,b)}$, where $a = 2+\vecalpha'\matsig^{-1}\vecalpha$ and $b = \delta(\vecx,\vecmu\mid\matsig)$.

Recall that the factor analysis model is given by \beq\vecN = \vecLambda\vecU+\vecepsilon,\eeq where $\vecLambda$, $\vecU$ and $\vecepsilon$ are defined as in Section~\ref{sec:introduction}. We can obtain the SAL factor analysis model via
\beq
\vecX = \vecmu+W\vecalpha+\sqrt{W}(\vecLambda\vecU+\vecepsilon).
\eeq
It follows that the marginal distribution of $\vecX$ is $\text{SAL}(\vecalpha,\vecLambda\vecLambda'+\vecPsi,\vecmu)$, and the mixture of SAL factor analyzers model has density 
\beq
f(\vecx\mid\varthet)=\gsum{\pi_g\xi(\vecx\mid\vecalpha_g,\vecLambda_g\vecLambda_g'+\vecPsi_g,\vecmu_g)}.
\eeq
Setting $\vecPsi_g=\omega_g\del_g$, we obtain a mixture of modified SAL factor analyzers with density 
\beq
f(\vecx\mid\varthet)=\gsum{\pi_g\xi(\vecx\mid\vecalpha_g,\vecLambda_g\vecLambda_g'+\omega_g\del_g,\vecmu_g)},
\eeq
and we can proceed in an analogous fashion to \cite{mcnicholas10d} to obtain a family of 12 parsimonious shifted asymmetric Laplace mixtures (PSALM; Table~\ref{PSALM}). This nomenclature for the PSALM family is analogous to that for the PGMM family except that the constrains are on component scale matrices rather than component covariance matrices. Note that the most general member of the PSALM family (UUUU), i.e., the mixture of (modified) SAL factor analyzers model, has component covariance matrix $\matsig_g+\vecalpha_g\vecalpha_g'$.
\begin{table*}[!ht]
\centering
\caption{Nomenclature, component scale structure, and number of free scale parameters for each member of the PSALM family.}\label{PSALM}
{\small\begin{tabular*}{1.02\textwidth}{@{\extracolsep{\fill}}ccccrr}\hline
\multicolumn{4}{c}{PSALM Nomenclature} &&\\
\cline{1-4}
$\load_g=\load$ & $\del_g=\del$ & $\omega_g=\omega$ & $\del_g=\ident_p$ & Cmp.\ Scale Matrix & Number of Free Scale Parameters\\\hline
C&C&C&C & $\matsig_g=\load\load'+\omega\ident_p$ &$[pq-q(q-1)/2]+1$\\
C&C&U&C & $\matsig_g=\load\load'+\omega_g\ident_p$ & $[pq-q(q-1)/2]+G$\\
U&C&C&C & $\matsig_g=\load_g\load_g'+\omega\ident_p$ & $G[pq-q(q-1)/2]+1$\\
U&C&U&C & $\matsig_g=\load_g\load_g'+\omega_g\ident_p$& $G[pq-q(q-1)/2]+G$\\
C&C&C&U &$\matsig_g=\load\load'+\omega\del$ & $[pq-q(q-1)/2]+p$\\
C&C&U&U & $\matsig_g=\load\load'+\omega_g\del$ & $[pq-q(q-1)/2]+[G+(p-1)]$\\
U&C&C&U & $\matsig_g=\load_g\load_g'+\omega\del$ & $G[pq-q(q-1)/2]+p$\\
U&C&U&U & $\matsig_g=\load_g\load_g'+\omega\del_g$ & $G[pq-q(q-1)/2]+[G+(p-1)]$\\
C&U&C&U & $\matsig_g=\load\load'+\omega\del_g$ & $[pq-q(q-1)/2]+[1+G(p-1)]$\\
C&U&U&U & $\matsig_g=\load\load'+\omega_g\del_g$ & $[pq-q(q-1)/2]+Gp$\\
U&U&C&U & $\matsig_g=\load_g\load_g'+\omega\del_g$ & $G[pq-q(q-1)/2]+[1+G(p-1)]$\\
U&U&U&U & $\matsig_g=\load_g\load_g'+\omega_g\del_g$ & $G[pq-q(q-1)/2]+Gp$\\
\hline
\end{tabular*}}
\end{table*}

\subsection{Parameter Estimation}

\subsubsection{The Alternating Expectation-Conditional Maximization Algorithm}

The expectation-maximization (EM) algorithm \citep{dempster77} is an iterative procedure used for estimating the maximum likelihood values of model parameters in the presence of missing or incomplete data. It is based on the complete-data vector, i.e., the unobserved together with the observed data, and iterates between two steps: an expectation (E-) step, where the expected value of the complete-data log-likelihood, $\mathcal{Q}$, is calculated, and a maximization (M-) step, where $\mathcal{Q}$ is maximized with respect to the model parameters. 

The expectation-conditional maximization (ECM) algorithm \citep{meng93} is a variant of the EM algorithm that replaces the M-step of the EM algorithm by a number of computationally efficient conditional maximization (CM-) steps. The alternating ECM (AECM) algorithm \citep{meng97} allows for the specification of different complete-data at each stage of the algorithm. 

\subsubsection{Deterministic Annealing}\label{sec:DA}

The deterministic annealing algorithm \citep{zhou09} is a modified EM algorithm that transforms the likelihood surface to enhance the chances of finding the dominant mode. In many model-based clustering approaches, e.g., the GPGM family, there is only one source of missing data, the component membership labels. For observed data $\vecx_1,\ldots,\vecx_n$, we denote these labels $\vecz_1,\ldots,\vecz_n$, where $\vecz_i=(z_{i1},\ldots,z_{iG})$ with 
\begin{equation*} z_{ig} = \left\{
\begin{array}{rl} 1 & \text{if $\mathbf{x}_i$ belongs to component } g,\\
0 & \text{otherwise}, \end{array} \right.
\end{equation*}
for $i=1,\dots,n$ and $g=1,\dots,G$. 

Because the random variable $\vecZ_i\mid\vecx_i$ follows the multinomial distribution with a single trial, we have
\beq
P(Z_{ig} = 1\mid\vecx_i) = \frac{\pi_gf_g(\vecx\mid\vectheta_g)}{\sum_{h=1}^G\pi_hf_h(\vecx\mid\vectheta_h)}.
\eeq
In the $E$-step of a standard EM algorithm, the component membership are estimated via
\beq
\hat{z}_{ig}\colonequals\mathbb{E}\left[Z_{ig}\mid\vecx_i\right]=\frac{\pi_gf_g(\vecx\mid\vectheta_g)}{\sum_{h=1}^G\pi_hf_h(\vecx\mid\vectheta_h)}.
\eeq
In deterministic annealing, however, the $E$-step is modified such that
\be
\mathbb{E}\left[Z_{ig}\mid\vecx_i\right]=\frac{[\pi_gf_g(\vecx\mid\vectheta_g)]^v}{\sum_{h=1}^G[\pi_hf_h(\vecx\mid\vectheta_h]^v},
\ee
where $v\in[0,1]$ is an auxiliary parameter drawn from a sequence of user-specified length. Note that the deterministic annealing algorithm is otherwise identical to the standard EM algorithm, or a variant thereof, and as it progresses, $v$ increases from $0$ to $1$.

\subsubsection{A Two-Phase Approach}

Using an EM algorithm for parameter estimation of the parameters in our SAL mixtures creates computational issues when updating $\hat{\vecalpha}_g$ and $\hat{\mSigma}_g$. This issue is attributed to the update of $\hat{\vecmu}_g$ sometimes taking on the value of an observation~$\vecx_i$. Although this estimate is legitimate, it makes updating $\hat{\vecalpha}_g$ and $\hat{\mSigma}_g$ impossible; specifically, the problem is around calculating the value of $K_{\nu}(u)$ and so the expected value $\mathbb{E}[W_{ig}^{-1} \mid \vecx_i,z_{ig}=1]$. To overcome this problem we propose a two-stage parameter estimation procedure where, in stage~1, we use a modified deterministic annealing algorithm, and in stage~2, an AECM algorithm. 

Because we now have three sources of missing data in each PSALM model, i.e., the latent $w_{ig}$, the component labels $\vecz_i$, and the latent factors $\vecu_{ig}$, the AECM algorithm is used for parameter estimation.
%
%
From Section~\ref{sec:model}, the complete-data log-likelihood can be written
\be\begin{split}\label{CDL}
l_{\text{c}}(\pi_g,\vecmu_g,\vecLambda_g,\omega_g,\del_g,\vecalpha_g)
&=\lsum\gsum{\big[\log{\pi_g}} + \log h\left(w_{ig}\right)\\
&\qquad\qquad+ \log\phi\left(\vecx_i\mid\vecmu_g+w_{ig}\vecalpha_g,w_{ig}\vecLambda_g\vecLambda_g'+w_{ig}\omega_g\del_g\right)\big]^{z_{ig}},
\end{split}\ee
where $\phi\left(\vecx_i\mid\vecmu_g+w_{ig}\vecalpha_g,w_{ig}\vecLambda_g\vecLambda_g'+w_{ig}\omega_g\del_g\right)$ is the density of a multivariate Gaussian distribution with mean $\vecmu_g+w_{ig}\vecalpha_g$ and covariance matrix $w_{ig}\vecLambda_g\vecLambda_g'+w_{ig}\omega_g\del_g$, and $h(w_{ig})$ is the density of an exponential distribution with rate 1, i.e., $h(z) = e^{-1}$, for real $z\geq 0$.

First, consider our modified deterministic annealing algorithm. At the fist stage, the complete-data consist of the data $\vecx_i$, the labels $z_{ig}$, and the latent $w_{ig}$, for $i=1,\ldots,n$ and $g=1,\ldots,G$. We update $\pi_g$, $\vecmu_g$, and $\vecalpha_g$, for $g=1,\ldots,G$, and use the following expected values:
%
\beq\begin{split}
&\mathbb{E}\left[Z_{ig}\mid\vecx_i\right]=\frac{[\pi_g\xi\left(\vecx_i\mid\vecalpha_g, \matsig_g,\vecmu_g\right)]^v}{\sum_{h=1}^G[\pi_h\xi\left(\vecx_i\mid\vecalpha_h, \matsig_h,\vecmu_h\right)]^v}\equalscolon\hat{z}^*_{ig},\\
&\Ewg= \sqrt{\frac{b_{ig}}{a_{g}}} R_{\nu}\left( \sqrt{a_{g} b_{ig}}\right)\equalscolon E_{1ig},\\
& \Ewinvg=
\sqrt{\frac{a_{g}}{\psi+b_{ig}}} R_{\nu}\left( \sqrt{a_{g}(\psi+b_{ig})}\right)-\frac{2\nu}{\psi+b_{ig}}\equalscolon E^*_{2ig},
\end{split}\eeq
where $v$ is defined in Section~\ref{sec:DA}, $a_g\colonequals2+\vecalpha_g'\matsig_g^{-1}\vecalpha_g$, $b_{ig}\colonequals\delta(\vecx_i,\vecmu_g~|~\matsig_g)$, and we introduce the parameter $\psi$ to prevent $\vecmu_g$ from becoming equal to any observation $\vecx_i$. The introduction of $\psi$ is motivated by the limiting case of the generalized hyperbolic density discussed in Section~\ref{sec:GHD}. Moreover, it is because of the presence of $\psi$ that we refer to this deterministic algorithm as ``modified''. The updates for the mixing proportions are given by $\hat{\pi}_g = {n_g}/{n}$, and the updates for the the skewness and shift parameters are given by 
\beq\label{eq:mle1}
\hat{\vecalpha}_g = \frac{  (\lsum \zig^* E^*_{2ig})(\lssum\zjg^*\vecx_j) 
-n_g\lsum \zig^* E^*_{2ig}\vecx_i }{ (\lsum \zig^* E_{1ig} )(\lssum\zjg^* E^*_{2jg}) - n_g^2 },
\eeq
and
\beq\label{eq:mle2}
\hat{\vecmu}_g  = \frac{ (\lsum\zig^*  E_{1ig})( \lssum\zjg^* E^*_{2jg}\vecx_j) - n_g\lsum\zig^*\vecx_i }{ (\lsum \zig^* E_{1ig})(\lssum\zjg^* E^*_{2jg})-n_g^2},
\eeq
respectively, where $n_g = \lsum\hat{z}^*_{ig}$.

At the second stage of our modified deterministic annealing algorithm, the complete-data include the same constituents as at the first stage plus the latent factors $\vecu_{ig}$, for $i=1,\ldots,n$ and $g=1,\ldots,G$. We update the factors loadings $\vecLambda_g$, the constant $\omega_g$, and the diagonal matrix $\del_g$. We need the same expected values from the first stage as well as:  
\beq\begin{split}
&\mathbb{E}[Z_{ig}W_{ig}^{-1}\vecU_{ig}\mid\vecx_i,w_{ig}^{-1}]=w_{ig}^{-1}\Beta_g(\vecx_i-\vecmu_g),\\
&\mathbb{E}[Z_{ig}W_{ig}^{-1}\vecU_{ig}\vecU_{ig}'\mid\vecx_i,w_{ig}^{-1}]=
\ident_q-\Beta_g\vecLambda_g+w_{ig}^{-1}\Beta_g(\vecx_i-\vecmu_g)(\vecx_i-\vecmu_g)'\Beta_g',
\end{split}\eeq
where $\Beta_g=\vecLambda_g'(\vecLambda_g\vecLambda_g'+\omega_g\del_g)^{-1}$. In each case, $w_{ig}^{-1}$ will be replaced by $E^*_{2ig}$. Note that these expected values are similar to those used by  \cite{mcnicholas08} and others.
The updates for $\vecLambda_g$, $\omega_g$, and $\del_g$ will depend on which PSALM family member is under consideration. Consider, for example, the mixture of SAL factor analyzers model (i.e., UUUU). In this case, the updates are 
\beq\begin{split}
&\hat\vecLambda_g=\sampcov_g\Beta_g'(\ident_q - \Beta_g\vecLambda_g +\Beta_g\sampcov_g\Beta_g')^{-1},\\
&\hat\omega_g=|\hat\vecPsi_g|^{1/p},\\ 
&\hat\del_g=\hat\vecPsi_g/|\hat\vecPsi_g|^{1/p},
\end{split}\eeq
where $\hat\vecPsi_g=\diag\{\sampcov_g-\vecLambda_g\hat\Beta_g \sampcov_g\}$ and
\beq\begin{split} 
\mathbf{S}_g&=\frac{1}{n_g}\lsum\zig^* E^*_{2ig} \left(\vecx_i-{\vecmu}_g\right)\left(\vecx_i-{\vecmu}_g\right)' 
- \hat{\vecalpha}_g\mathbf{r}_g' - \mathbf{r}_g\hat{\vecalpha}_g'
+ \frac{1}{n_g}\hat{\vecalpha}_g\hat{\vecalpha}_g'\lsum\zig E_{1ig},
\end{split} \eeq 
where $\mathbf{r}_g \colonequals ({1}/{n_g})\lsum\zig\left(\vecx_i-\hat{\vecmu}_g\right)$.
Updates for the other models (Table~\ref{PSALM}) are analogous to those given in \cite{mcnicholas08,mcnicholas10d}. Our modified deterministic annealing algorithm starts at $v=0$ and stops at $v=1$, having iterated over a user-specified sequence (cf.\ Section~\ref{sec:DA}).



After our deterministic annealing algorithm is run, the second phase of our parameter estimation procedure is an AECM algorithm. This AECM algorithm is identical to our modified deterministic annealing algorithm, except that now
\be\label{eq:exp2}
\mathbb{E}\left[Z_{ig}\mid\vecx_i\right]=\frac{\pi_g\xi\left(\vecx_i\mid\vecalpha_g, \matsig_g,\vecmu_g\right)}{\sum_{h=1}^G\pi_h\xi\left(\vecx_i\mid\vecalpha_h, \matsig_h,\vecmu_h\right)}\equalscolon\zig,
\ee
and
\beq\begin{split}
 \Ewinvg &=\sqrt{\frac{a_{g}}{b_{ig}}} R_{\nu}\left( \sqrt{a_{g}(b_{ig})}\right) -\frac{2\nu}{b_{ig}}
\equalscolon E_{2ig},
\end{split}\eeq
and on the first CM step we update only $\hat{\pi}_g$ and $\hat{\vecalpha}_g$, for $g=1,\ldots,G$. These updates, as well as the updates for $\hat{\vecLambda}_g$, $\hat{\omega}_g$, and $\hat{\del}_g$, are the same as in our modified deterministic annealing algorithm but with $\zig^*$ and $E^*_{2ig}$ replaced by $\zig$ and $E_{2ig}$, respectively. Accordingly, we now have
\beq\begin{split} 
\mathbf{S}_g&=\frac{1}{n_g}\lsum\zig^* E^*_{2ig} \left(\vecx_i-{\vecmu}_g\right)\left(\vecx_i-{\vecmu}_g\right)'
- \hat{\vecalpha}_g\mathbf{r}_g' - \mathbf{r}_g\hat{\vecalpha}_g'
+ \frac{1}{n_g}\hat{\vecalpha}_g\hat{\vecalpha}_g'\lsum\zig E_{1ig}
\end{split} \eeq 
in the updates for $\hat{\vecLambda}_g$, $\hat{\omega}_g$, and $\hat{\del}_g$. Convergence of our AECM algorithms is discussed in Section~\ref{sec:convergence}.

After convergence, clustering results are reported based on maximum \textit{a~posteriori} (MAP) classification values. That is, $\text{MAP}\{\zig\}=1$ if max$_h\{\hat{z}_{ih}\}$ occurs in component $h=g$, and $\text{MAP}\{\zig\}=0$ otherwise. In other words, each observation is assigned to the component to which it has the highest 
\textit{a~posteriori} of membership.

\section{Computational Considerations}\label{sec:CC}

\subsection{Initialization}

When initializing the component membership labels $\zig$, we use random starting values in both model-based clustering and classification applications. Initial values for $\pi_g$ and $\vecmu_g$ follow directly. Following \cite{mcnicholas08}, we initialize the parameters $\load_g$, $\omega_g$, and $\del_g$ based on the eigen-decomposition of $\sampcov_g$. Frist, the $\sampcov_g$ are computed based on the initial values of the $\zig$. The initial values of the elements of $\load_g$ are set as
\beq\lambda_{ij} = \sqrt{d_{j}}\rho_{ij},\eeq
where $d_{j}$ is the $j$th largest eigenvector of $\sampcov_g$ and
$\rho_{ij}$ is the $i$th element of the $j$th largest eigenvector
of $\sampcov_g$, for $i =1,2,\ldots,p$ and $j = 1,2,\ldots,q$. The $\omega_g\del_g$ are then initialized based on 
\beq\begin{split}
&\hat\omega_g=|\diag\{\sampcov_g - \load_g\load_g'\}|^{1/p},\\ 
&\hat\del_g=\frac{1}{|\diag\{\sampcov_g - \load_g\load_g'\}|^{1/p}}\diag\{\sampcov_g - \load_g\load_g'\}.\\
\end{split}\eeq

\subsection{Woodbury Identity}

To avoid inverting any non-diagonal $p\times p$ matrices we make use of the Woodbury Identity \citep{woodbury50}, which states that
\beq
(\vecA+\vecU\vecC\vecV)^{-1}=\vecA^{-1}-\vecA^{-1}\vecU(\vecC^{-1}+\vecV\vecA^{-1}\vecU)^{-1}\vecV\vecA^{-1},
\eeq
where $\vecA$ is an $m\times m$ matrix, $\vecU$ is an $m\times k$ matrix, $\vecC$ is an $k\times k$ matrix, and $\vecV$ is an $k\times m$ matrix. Setting $\vecU = \vecLambda$, $\vecV = \vecLambda'$, $\vecA=\omega\del$, and $\vecC=\ident_q$ we can write 
\beq\begin{split}
&(\omega\del+\vecLambda\vecLambda)^{-1}=\\
&\qquad(\omega\del)^{-1}-(\omega\del)^{-1}\vecLambda(\ident_q+\vecLambda'(\omega\del)^{-1}\vecLambda)^{-1}\vecLambda'(\omega\del)^{-1}.
\end{split}\eeq
Following from this we can compute the determinant using,
\beq
|\omega\del+\vecLambda\vecLambda'|=|\omega\del|/|\ident_q+\vecLambda'(\vecLambda\vecLambda'+\omega\del)^{-1}\vecLambda|.
\eeq
These identities provide a major computational advantage as $p$ grows, because $q\ll p$, and have been used by many authors including \cite{mclachlan00a}, \cite{mcnicholas08}, \cite{andrews11a}, \cite{andrews11b}, and \cite{murray13}.

\subsection{Model-Based Classification}
In a true clustering scenario, no observations have known labels. However, if some proportion of the group memberships are known, we can use them to help estimate the component labels for the remaining $n-k$ observations. One approach to do this is `model-based classification', which is a semi-supervised version of model-based clustering \citep[e.g.,][]{dean06,mcnicholas10c,andrews11d}. 
Model-based classification is preformed within a joint likelihood framework. Without loss of generality, we order the observations $\vecx_1,\ldots,\vecx_k, \vecx_{k+1},\ldots,\vecx_n$ so that it is the first~$k$ observations that are labelled. The PSALM model-based classification likelihood is given by
\begin{equation}\begin{split}\label{joint}
\mathcal{L}\left(\vecx_1,\ldots,\vecx_n,\vecz_1,\ldots,\vecz_n\mid\varthet\right)=&
\prod_{i=1}^k\prod_{g=1}^G{\lbrack\pi_g\xi\left(\vecx_i\mid\vecalpha_g,\vecLambda_g\vecLambda_g'+\omega_g\del_g,\vecmu_g\right)\rbrack^{z_{ig}}}\\&\qquad
\times\prod_{j=k+1}^{n}\sum_{h=1}^H{\pi_h\xi\left(\vecx_j\mid\vecalpha_h, \vecLambda_h\vecLambda_h'+\omega_h\del_h,\vecmu_h\right)},\end{split}\end{equation}
where $H\geq G$. Parameter estimation is analogous to model-based clustering (cf. Section~\ref{sec:meth}).

\subsection{Convergence}\label{sec:convergence}

We use a criterion based on Aitken's acceleration to determine whether our AECM algorithm has converged. The algorithm can be considered to have converged when
$l_{\infty}^{(t+1)} - l^{(t+1)}<\epsilon$,
where $l^{(t+1)}$ is the log-likelihood at iteration $t+1$ and \beq l_{\infty}^{(t+1)} = l^{(t)} + \frac{l^{(t+1)}-l^{(t)}}{1 - a^{(t)}}\eeq is an asymptotic estimate of the log-likelihood at iteration $t+1$ \cite[cf.][]{bohning94}.
At iteration~$t$, the Aitken acceleration is
\beq a^{(t)}=\frac{l^{(t+1)} - l^{(t)}}{l^{(t)} - l^{(t-1)}},\eeq
where $l^{(t+1)}$, $l^{(t)}$, and $l^{(t-1)}$ are the log-likelihood values at iterations $t+1$, $t$, and $t-1$, respectively. 
We stop our AECM algorithm when $l_{\infty}^{(t+1)} - l^{(t)}<\epsilon$ \citep[cf.][]{lindsay95}; in our applications we use $\epsilon = 0.01$.

\subsection{Model Selection}

\subsubsection{Bayesian Information Criterion}
The Bayesian information criterion \citep[BIC;][]{schwarz78} is a popular tool used for mixture model selection \citep{dasgupta98,leroux92,keribin00}. The BIC is given by \beq\text{BIC}=2l(\vecx\mid\hat{\varthet})-\rho\log{n},\eeq where $l(\vecx\mid\hat{\varthet})$ is the maximized log-likelihood, $\hat{\varthet}$ is the maximum likelihood estimate of $\varthet$, $\rho$ is the number of free parameters in the model, and $n$ is the number of observations. \cite{lopes04} illustrate the use of the BIC for selecting the number of factors in a factor analysis model, and the BIC has often been used for model selection in families of mixture models with latent factors \citep[e.g.][]{mcnicholas08,mcnicholas10d,andrews11a,andrews11b,murray13}.

\subsubsection{Integrated Completed Likelihood}
The integrated completed likelihood \citep[ICL;][]{biernacki00} penalizes the BIC for uncertainty in classification, which some argue makes it more for suitable model-based clustering and classification applications. The ICL is given by
\beq\text{ICL} \approx \text{BIC} + \sum_{i=1}^n\sum_{g=1}^G{\text{MAP}\{\zig\}\log{\zig}},\eeq
where $\sum_{i=1}^n\sum_{g=1}^G{\text{MAP}\{\zig\}\log{\zig}}$ is the estimated mean entropy and reflects the uncertainty in the classification of observations into components. \cite{franczak13} used the ICL for model selection in the case of SAL mixtures. 

Herein, we compare the performance of the ICL and BIC for model selection for the PSALM family. Note that, in our applications, the model selection problem is three-fold in that we must choose the member of the family (Table~\ref{PSALM}) as well as the numbers of components and factors, respectively.

\subsection{Performance Assessment}

We treat each of our applications (Section~\ref{sec:Crabs}) as genuine clustering or classification examples. However, the labels are actually known in each case and this allows us to investigate classification performance. We use the adjusted Rand index \cite[ARI;][]{hubert85} to assess the classification performance of our PSALM models. The Rand index \citep{rand71} is given by
\beq
\frac{\text{number of agreements}}{\text{number of agreements + number of disagreements}}
\eeq
and is used to compare two partitions, e.g., true and predicted classification. Unfortunately, the Rand Index has an expected value that is greater than 0 under random classification and so interpretation of smaller values is difficult. The ARI corrects the Rand index for chance, and it has an expected value of~0 under random classification and a value of 1 under perfect class agreement.

\section{Applications}\label{sec:Apps}

\subsection{Leptograpsus Crabs}\label{sec:Crabs}

\cite{campbell74} give data on 200 crabs of the species \textit{Leptograpsus variegatus} collected at Fremantle, Western Australia. The data are available in the \textsf{R} package \texttt{MASS} and contain 5 morphological measurements: frontal lobe size (mm), rear width (mm), carapace length (mm), carapace width (mm) and body depth (mm). The crabs are of two genders and two colours (blue and orange). In this section, 
we consider the principal components of the crabs data, and in Section~\ref{sec:MBClass}, we consider the full data set in both model-based clustering and classification scenarios.

\begin{figure*}[!htb]
\centering%
\includegraphics[height=2.85in,width=6in]{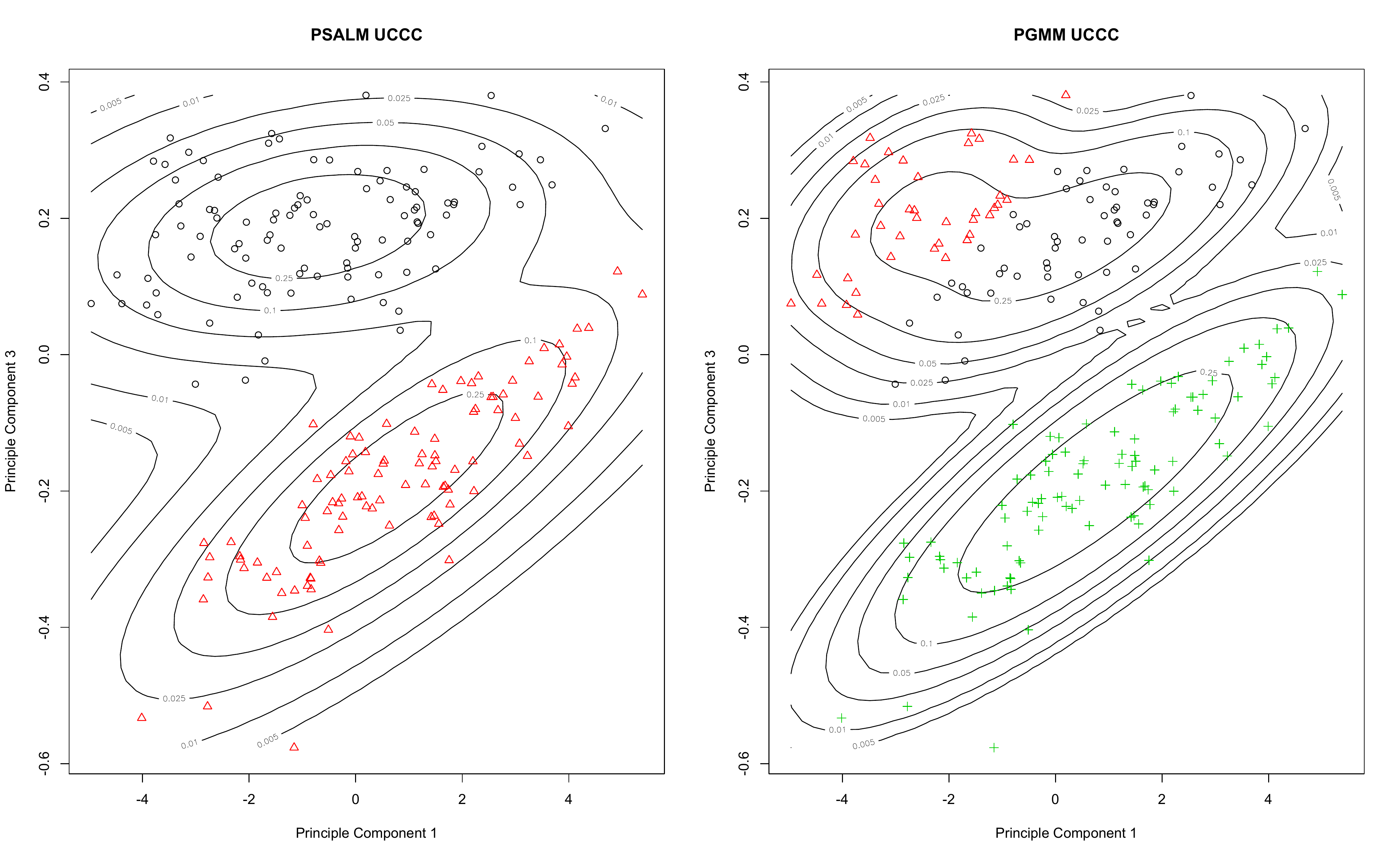}
\caption{The first and third principal components of the crabs data with contours reprinting the fit of the chosen PSALM and PGMM models, respectively.}\label{Crabs1}
\end{figure*}

As one might expect, the variables in the crabs data are highly correlated with one another.
Using principal component analysis \citep[PCA;][]{pearson01,hotelling33} on the crabs data shows that $95\%$ of the variation in the data can be explained by the first three principal components. Interestingly, plotting the first and third principal components against each other results in two equally sized groups, where each group represents one gender (Figure~\ref{Crabs1}). 
The PSALM and EPGMM families were fit to the first three principal components for $G=1,\dots,9$ groups and $q=1$ latent factor. The best PSALM model, as chosen by both the BIC ($-812.8928$) and ICL ($-815.8221$), was the $G=2$ component UCCC model, which gave perfect classification with respect to gender ($\text{ARI} = 1.00$). The best PGMM also had the UCCC covariance structure ($\text{BIC} = \text{ICL} = -730.7081$) but used $G=3$ components to fit the data ($\text{ARI} = 0.756$). Figure~\ref{Crabs1} shows the resulting MAP classifications as well as contours for the fitted PSALM and PGMM models, respectively. Notably, the classifications for the chosen PGMM model can be made equal to those for the chosen PSALM model by merging components.

\subsection{Gaussian Cluster Merging \& Swiss Bank Notes}

The results for the chosen PGMM model in Section~\ref{sec:Crabs} raise an interesting point, i.e., that merging Gaussian mixture components can sometimes give identical classification performance to a non-Gaussian mixture (cf.\ Figure~\ref{Crabs1}). \cite{baudry10} and \cite{hennig10} discuss methods for merging components. Now, we will look at another example to reinforce the point that the performance of non-Gaussian approaches can sometimes be matched by Gaussian mixtures followed by merging. 

Consider the Swiss bank notes data, which are available in the \textsf{R} package \texttt{gclus} \citep{hurley04}. The bank notes data are composed of 200 swiss bills, of which 100 are counterfeit and 100 are legitimate. There are six physical measurements available for each bill. The PSALM and PGMM families were fit to these data for $G=1,\dots,9$ components and $q=1,2,3$ latent factors. The classification results (Table~\ref{ClustBank}) show that, like the analysis of the crabs principal components, the classification performance of the chosen PGMM model (CUCU) can be made equal to that of the chosen PSALM model (CCCU) by merging components.
\begin{table}[!ht]
\centering%
\caption{Clustering results for the chosen PSALM and PGMM for the swiss bank notes data. The bill types are cross-tabulated against our predicted classifications (A, B) in each case.}\label{ClustBank}
\begin{tabular*}{1\textwidth}{@{\extracolsep{\fill}}ccccccccc}
\hline
& \multicolumn{2}{c}{PSALM} && \multicolumn{4}{c}{PGMM} \\
\cline{2-3}\cline{5-8}
& A & B && A & B & C & D \\
\hline
Counterfeit & $99$ & $1$ && $21$ & $0$ & $1$ & $78$ \\
Legitimate & $0$ & $100$ && $0$ & $84$ & $16$ & $0$ \\
\hline
\end{tabular*} 
\end{table}

\subsection{Yeast Data}

\cite{nakai91, nakai92} discuss the development and classification results for the cellular localization sites of 1,484 yeast proteins. This data set is available in the UCI machine learning repository and in our analysis we consider only three variables: McGeoch's method for signal sequence recognition ({\tt mcg}), the score of the ALOM membrane spanning region prediction program ({\tt alm}), and the score of discriminant analysis of the amino acid content of vacuolar and extracellular proteins ({\tt vac}). We attempt to distinguish between the two localization sites, CYT (cytosolic or cytoskeletal) and ME3 (membrane protein, no N-terminal signal).

\begin{figure}[!htb]
{\includegraphics[height=2.80in,width=8.5cm]{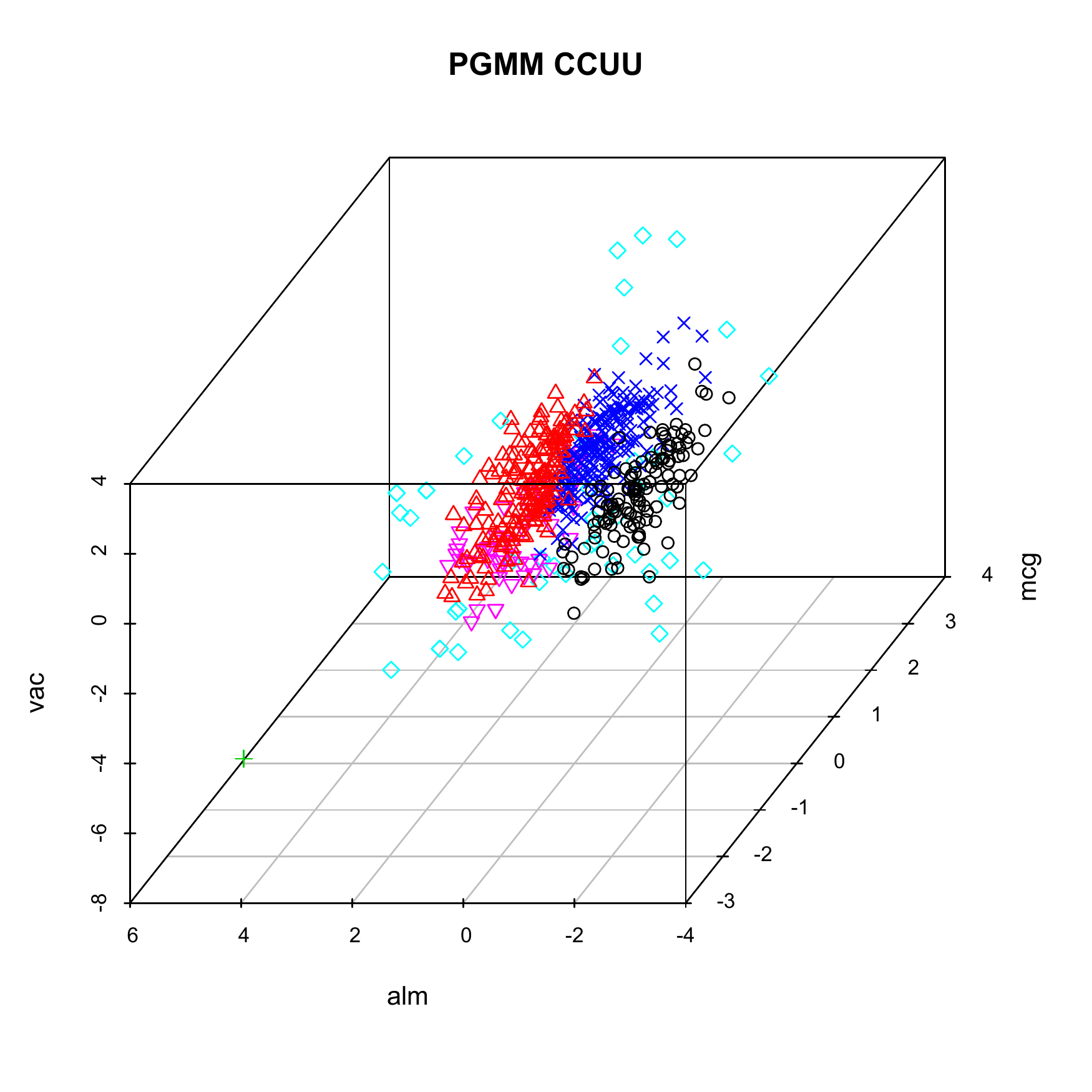}}
{\includegraphics[height=2.80in,width=8.5cm]{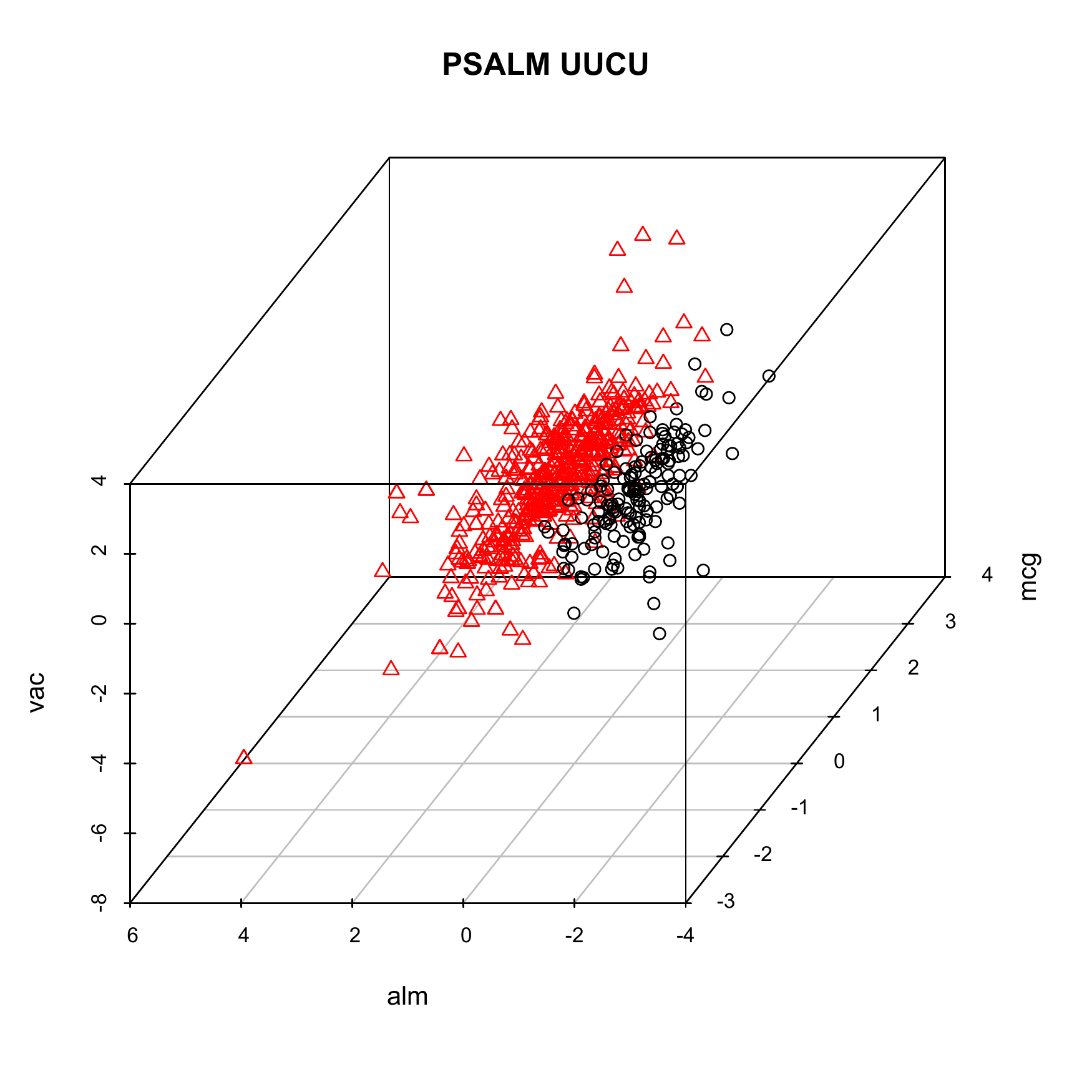}}
{\includegraphics[height=2.80in,width=8.5cm]{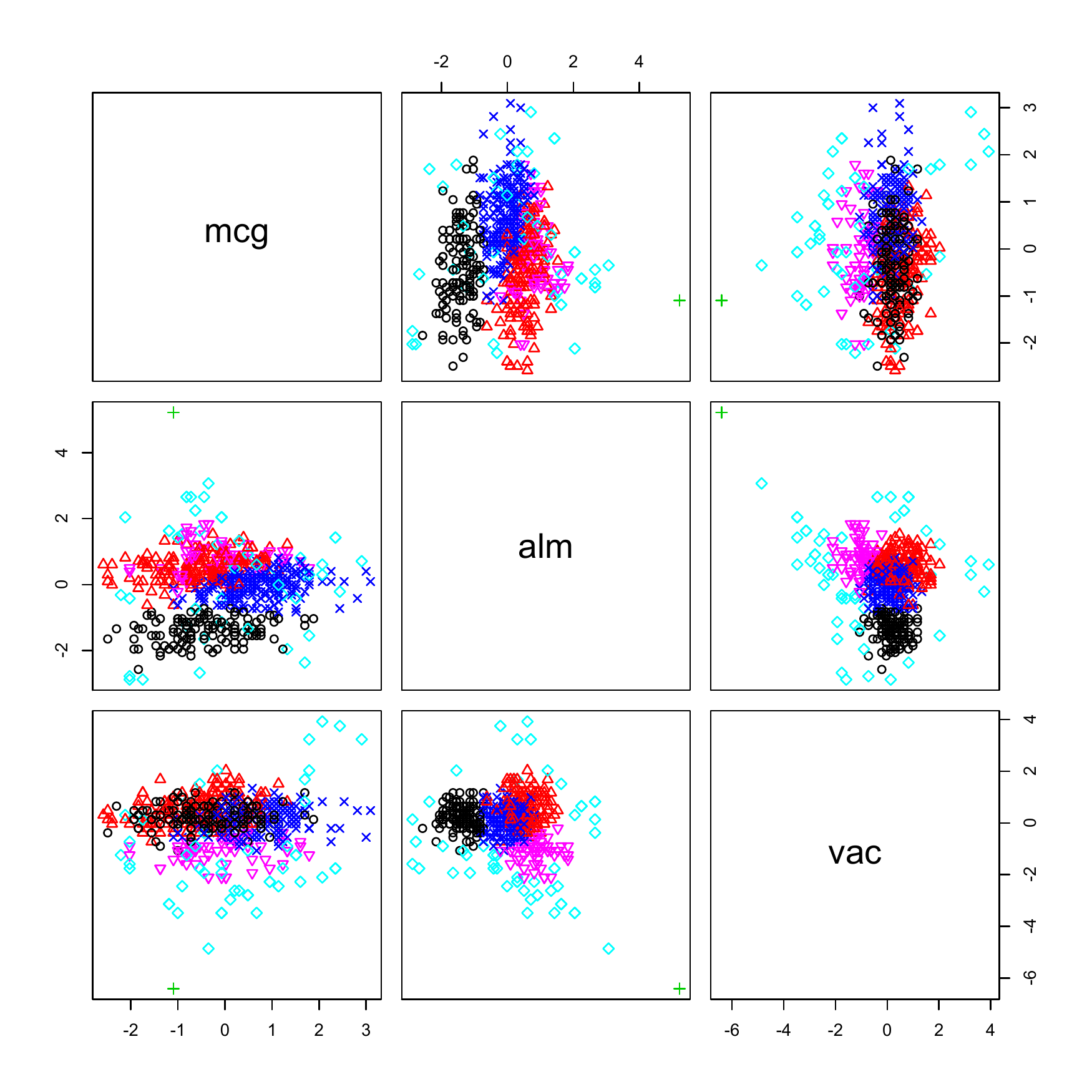}}
{\includegraphics[height=2.80in,width=8.5cm]{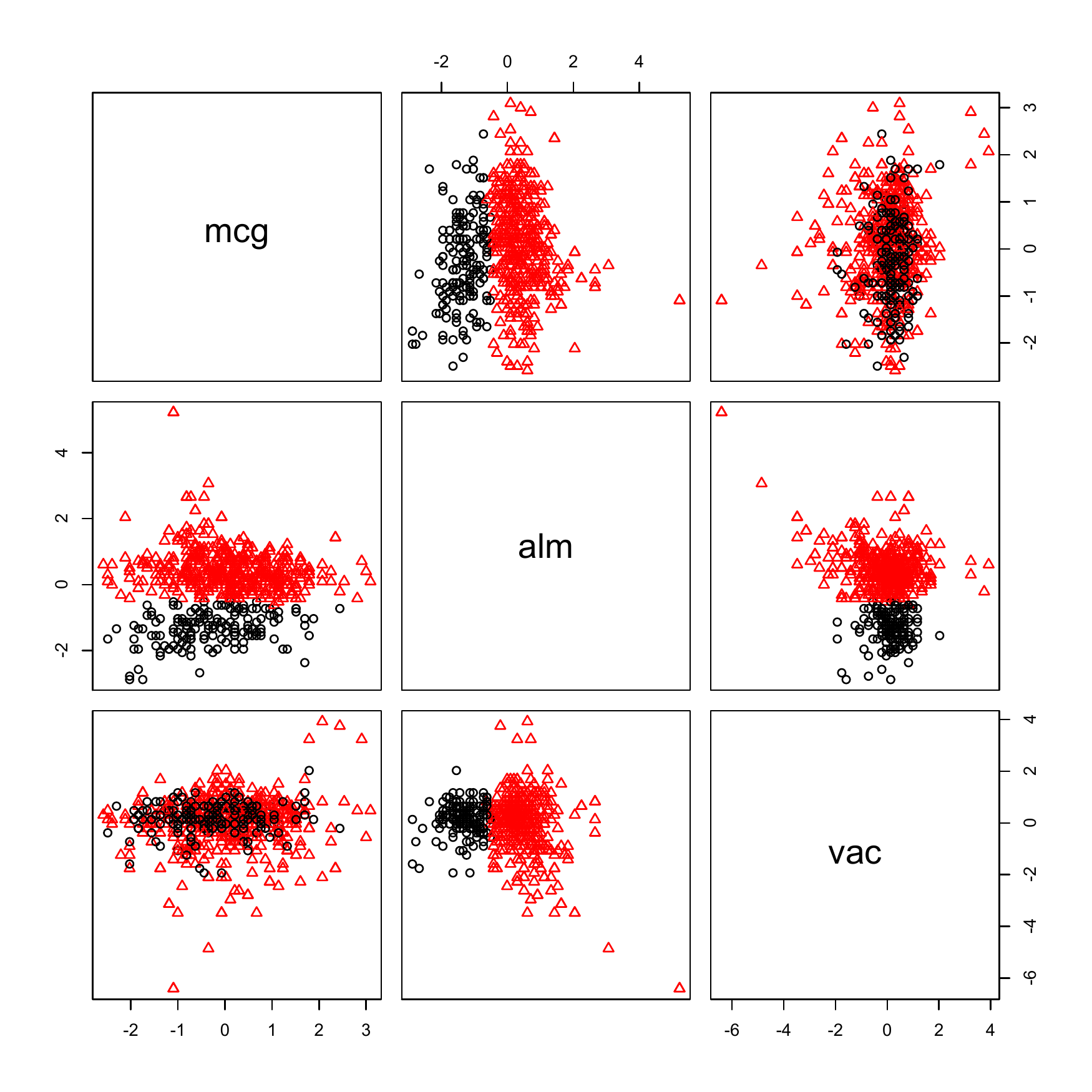}}
\caption{The yeast data with predicted group memberships highlighted for the chosen PSALM and PGMM models.}\label{YeastCol2}
\end{figure}

The PSALM and PGMM families were fitted to these yeast data for $G = 1,\dots,9$ components and $q=1$ latent factor. The chosen PGMM model is a $G=6$ component CCUU model $(\text{BIC} = \text{ICL} = -4993.137, \text{ARI} = 0.22)$. For the PSALM family, the ICL ($-5275.705$) selects a $G=2$ component CUUU model ($\text{ARI} = 0.86$), and the BIC ($-5220.763$) chooses a $G=2$ component UUCU model ($\text{ARI} = 0.83$). The classification results for the best fitting Gaussian and SAL mixtures, as chosen by BIC, are given in Table~\ref{ClustYeast}.
From this table and Figure~\ref{YeastCol2}, we can see that merging components could improve the PGMM solution. However, even under the optimal merging scenario, i.e., combining PGMM components B through F (cf.\ Table~\ref{ClustYeast}), results in classification performance ($\text{ARI}=0.705$) that is still not as good the chosen PSALM model.
\begin{table*}[!htb]
\centering%
\caption{Clustering results for the chosen PSALM and PGMM for the yeast data. The cellular localization sites are cross-tabulated against our predicted classifications (A, B, C, D, E, F) in each case.}\label{ClustYeast}
\begin{tabular*}{0.99\textwidth}{@{\extracolsep{\fill}}cccccccccc}
\hline
& \multicolumn{2}{c}{PSALM UUCU} && \multicolumn{6}{c}{PGMM CCUU} \\
\cline{2-3}\cline{5-10}
& A & B && A & B & C & D & E & F \\
\hline
CYT & $454$ & $9$ && $4$ & $170$ & $2$ & $178$ & $40$ & $69$\\
ME3 & $13$ & $150$ && $123$ & $5$ & $0$ & $22$ & $12$ & $1$\\
\hline
\end{tabular*} 
\end{table*} 

\subsection{Model-Based Classification}\label{sec:MBClass}

Thus far, our applications have focused on model based clustering. Now, we consider applying the PSALM and PGMM families in model-based classification scenarios to crabs and yeast data, respectively. In each case, we take 50 random subsets of $80\%$ of the labels to be known and we assume that we know the number of classes. Note that we consider the full $p=5$ variable, $G=4$ component crabs data set, which presents a much more challenging classification problem than using the principal components. The aggregate classification results are given in Tables~\ref{PGMMCrabs} and~\ref{ClassYeast}. In each case, PSALM clearly outperforms the PGMM family. 
\begin{table*}[!htb]
\centering%
\caption{Aggregated classification results for the PSALM CCCU (ARI = 0.853) and PGMM UCUU (ARI = 0.737) for the Leptograpsus crabs data. The colour and gender combinations are cross-tabulated against our predicted classifications (A, B,C,D) in each case.}\label{PGMMCrabs}
\begin{tabular*}{\textwidth}{@{\extracolsep{\fill}}cccccccccc}
\hline
& \multicolumn{4}{c}{PSALM} && \multicolumn{4}{c}{PGMM} \\
\cline{2-5}\cline{7-10}
& A & B & C & D && A & B & C & D \\
\hline
Blue Male & $461$ & $47$ & $0$ & $0$ && $256$ & $148$ & $0$ & $0$ \\
Orange Male & $32$ & $501$ & $0$ & $0$ && $0$ & $745$ & $0$ & $0$ \\
Blue Female & $0$ & $0$ & $463$ & $0$ && $0$ & $0$ & $205$ & $0$ \\
Orange Female & $0$ & $7$ & $29$ & $460$ && $0$ & $0$ & $98$ & $548$ \\
\hline
\end{tabular*} 
\end{table*} 
\begin{table}[!ht]
\centering%
\caption{Clustering results for the chosen PSALM and PGMM for the yeast data. The localization sites are cross-tabulated against our predicted classifications (A, B) in each case.}\label{ClassYeast}
\begin{tabular*}{1\textwidth}{@{\extracolsep{\fill}}cccccc}
\hline
& \multicolumn{2}{c}{PSALM CCCU} && \multicolumn{2}{c}{PGMM CCUC} \\
\cline{2-3}\cline{5-6}
& A & B && A & B \\
\hline
CYT & $4591$ & $48$ && $4450$ & $150$ \\
ME3 & $184$ & $1477$ && $250$ & $1450$ \\
\hline
\end{tabular*} 
\end{table} 





\section{Discussion}\label{sec:discuss}

We have extended the mixture of factor analyzers model using SAL mixtures. Based on the resulting mixture of (modified) SAL factor analyzers model, a new family of mixture models, i.e., PSALM, was developed. The PSALM models are well suited for the analysis of high-dimensional data because the covariance structure allows for $p$-dimensional data to be represented by $q$ latent factors where $q \ll p$; crucially, the number of covariance parameters is linear in data dimensionality for each member of the PSALM family.

A two-stage approach was taken to parameter estimation for members of the PSALM family, consisting of deterministic annealing followed by an AECM algorithm. The performance of the PSALM family was compared to the PGMM family in both model-based clustering and classification scenarios. In these applications, the BIC and the ICL were used for model selection. Our PSALM models gave similar or superior classification performance to the PGMM family and although merging could sometimes be used to bring the classification performance of the PGMM models up to that of the PSALM family, this was not always the case. 

Future work will include more efficient implementation of the PSALM family, developing SAL mixtures suitable for the analysis of longitudinal data \citep[cf.][]{mcnicholas10a,mcnicholas12}, and developing SAL analogues of the common factor analyzers model \citep{baek10,baek11,murray13b}. We note that the special case of the mixture of SAL factor analyzers model introduced herein with $G=1$ corresponds to the SAL factor analysis model, which is itself worthy of further study.

\section*{Acknowledgements}
This work was supported by an Ontario Graduate Scholarship, the University Research Chair in Computational Statistics, and an Early Researcher Award from the Ontario Ministry of Research and Innovation.


\end{document}